\documentclass[aps,prl,reprint,groupedaddress,showpacs]{revtex4-1}

\usepackage{graphicx}
\usepackage{amssymb}
\usepackage{amsmath}
\usepackage{natbib}
\usepackage{color}

\begin{document}

\title{ $\mathcal{PT}$-symmetric scattering in flow duct acoustics} 

\author{Yves Aur\'egan}\email{yves.auregan@univ-lemans.fr}
\affiliation{Laboratoire d'Acoustique de l'Universit\'e du Maine, UMR CNRS 6613
Av. O Messiaen, F-72085 LE MANS Cedex 9, France}
\author{Vincent Pagneux} \email{vincent.pagneux@univ-lemans.fr}
\affiliation{Laboratoire d'Acoustique de l'Universit\'e du Maine, UMR CNRS 6613
Av. O Messiaen, F-72085 LE MANS Cedex 9, France}

\begin{abstract}

We show  theoretically and experimentally that the propagation of an acoustic wave in an airflow duct going through a pair of diaphragms, 
{ with equivalent amount of mean-flow-induced effective gain and  loss}, 
displays all the features of a parity-time ($\mathcal{PT}$) symmetric system. 
Using a scattering matrix formalism, 
we observe experimentally the  properties which reflect the $\mathcal{PT}$-symmetry of the scattering acoustical system:  
the existence of a spontaneous symmetry breaking with symmetry-broken pairs of scattering eigenstates showing amplification and reduction, 
and the existence of points with unidirectional invisibility.

\end{abstract}


\maketitle 


Hydrodynamic instability theory shows that flow can provide energy to small perturbations
\cite{schmid,drazin}. 
If, in addition, these perturbations are compressible, then both acoustic wave propagation and energy exchange
with the flow are possible, leading e.g. to the classical whistling phenomena \cite{mohring83,goldstein,fabrikant}.
Thus, in the particular case of  flow duct acoustics, the wave can obviously be  convected  but it also experiences gain or loss due to 
interactions with the flow inhomogeneities \cite{howe}. 
Consequently, propagation of acoustic waves in ducts with flow is a natural Non-Hermitian system where
loss and gain are available.

Non-Hermitian systems, where energy conservation is broken, lead to dynamics governed by evolution equations with 
non-normal operators, where surprising phenomena can appear due to huge non-normality especially close to 
exceptional points \cite{trefethen,moiseyev,krejcirik}.
The particular case of $\mathcal{PT}$-symmetry, where gain and  loss are delicately balanced,
has attracted a lot of attention in the last two decades
\cite{bender1998,mostafa2002,bender2005,bender2007,guo2009,christo2010,Christodoulides2007,Regensburger2013,cerjan2016,christo2016}. 
It opens the possibility  to obtain purely real spectra from Non-Hermitian Hamiltonians, 
as well as a spontaneous symmetry breaking where real eigenvalues coalesce at an exceptional point to become complex conjugate pair.
From a scattering point of view, another type of spontaneous symmetry breaking for $\mathcal{PT}$-symmetric systems has been theoretically proposed \cite{stone2011}.
It corresponds to the transition of norm-preserving scattering eigenstates, with unimodular eigenvalues, to symmetry broken pairs 
of amplified and lossy scattering eigenstates, with associated pairs of scattering eigenvalues with 
inverse moduli \cite{stone2011,stone2012,stone2013,diakonos2014,zhu2016}.
 It is to be noticed that this type of symmetry breaking is still waiting to be observed experimentally \cite{lige2015,lige2016}.
 
Initiated in the domain of quantum mechanics, many works on $\mathcal{PT}$-symmetry 
have displayed several intriguing effects such as 
power oscillation \cite{Christodoulides2009,christo2010,Regensburger2012,koslov2015}, 
unidirectional transparency \cite{kottos2011,longhi2011,longhi2014}, 
single-mode laser \cite{zhang2014c,hoadei2014}, spectral singularity and Coherent Perfect Absorber (CPA)-Laser
 \cite{mostafazadeh2009,longhi2010,stone2011,zhang2014b,feng2016}
or enhanced sensitivity \cite{nori2016}.
A majority of the studies has been conducted in optics with
some attempts in acoustics where the difficulty to obtain gain has been recognized.
Actually, whilst losses can be easily  introduced \cite{vrg2016a,vrg2016b}, the gain for acoustic waves has until now been obtained owing
to active electric amplification \cite{ramezani2014,fleury2015,zhang2016,christensen2016}. 

\begin{figure}[h]
\includegraphics[width=\columnwidth]{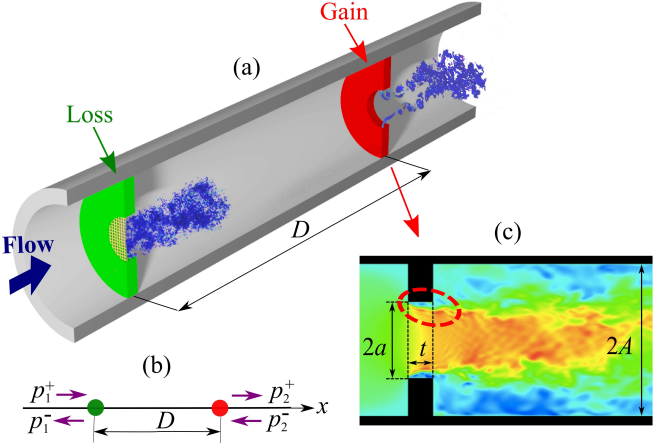}%
\caption{\label{fig_1} (a) Sketch of the acoustic $\mathcal{PT}$ symmetric system in an airflow duct. (b) Corresponding 1D model. (c) Numerical simulation with Large Eddies Simulation \cite{lacombe2013} of the flow in the diaphragm in the presence of  an acoustic wave. }%
\end{figure}

In this letter, we report the experimental realization of a purely mechanical scattering $\mathcal{PT}$-symmetric system 
for the propagation of acoustic waves in a waveguide. 
The loss and the gain are produced by 
two localized scattering units made of  diaphragms, one associated with loss and the other
associated with gain, see Fig. \ref{fig_1}(a).
In our experiments, the Mach number of the flow is small enough 
($\mathit{ M\kern-.20em a} \simeq 0.01$) such that the effect of convection on the sound wave can
be neglected, preserving the reciprocity property, and the only effect of the flow is located at the two diaphragms, 
characterized by  normalized complex impedances $C_1$ and $C_2$.
The balance of gain and loss is realized by finely tuning the flow rate and the geometry of each diaphragm, ensuring
a $\mathcal{PT}$-symmetric system that corresponds to  $C_1=C_2^*$ (note that the real part of the two impedances have to be equal to
get the parity symmetry).
Measurements of the scattering matrix components allow us to demonstrate unidirectional invisibility and to verify the $\mathcal{PT}$-symmetry properties.
Besides, by changing the distance between the scatterers,  the spontaneous symmetry breaking of the scattering matrix is observed with the transition from
 exact-$\mathcal{PT}$-symmetric  phase to  $\mathcal{PT}$-broken phase.
 In the broken phase, with the experimental gain available, the scattering eigenstates can be simultaneously fourfold amplified or reduced, 
 and we show that this effect might be enhanced by considering a finite periodic collection of the set of two diaphragms, leading to CPA-Laser points.

{\it System description and 1D model}.---
The description of the set-up is shown in Fig. \ref{fig_1}. 
We consider an acoustic waveguide where only plane waves can propagate ($k A < 1.841$ \cite{pierce}, where $A$ is the tube radius, $k=\omega/c_0$ is the wavenumber, $\omega$ is the frequency and $c_0$ is the sound velocity). The propagation for the acoustic pressure $p$ is then governed by the 1D Helmholtz equation. 
Two diaphragms are inserted into the tube and are separated with a distance $D$ (Fig. \ref{fig_1}(a)). 
As their thicknesses $t$ are small ($k t \ll 1$), the acoustic velocity is conserved while the pressure jumps between the two sides of the discontinuities. Thus the propagation is governed by
\begin{equation}
p'' \: + \: k^2 \:  p = 0,   
\end{equation}
 with the point scatterer jump conditions at the diaphragms: 
\begin{eqnarray*}
\Big \lbrack \;  p' \; \Big \rbrack_{x=\pm D/2}&=& 0 ,\\
\Big \lbrack \;  p \; \Big\rbrack _{x=- D/2}= \frac{C_1}{k} \, p'              \:\:\:\:  &\text{and}& \:\:\:\:        
\Big \lbrack \;  p \; \Big\rbrack _{x= D/2} = \frac{C_2}{k} \, p'.      
\end{eqnarray*}
where prime is the derivative  with respect to $x$. 
The real part of the dimensionless parameters $C_{1,2}$ is associated to reactive effects while its imaginary part is linked to the dissipative or gain effects. 
We have thus a very simple 1D reciprocal wave model with two point scatterers at $x=\pm D/2$ (Fig. \ref{fig_1}(b)).
The effect of the flow on acoustic propagation is only and entirely contained in the complex impedances
$C_1$ and $C_2$ that reflect the mean-flow-induced effective gain and loss.

The system is $\mathcal{PT}$-symmetric if and only if the two impedances are complex conjugated: $C_2=C_1^*$ \cite{mostafazadeh2010}. 
 With the $\exp(-\mathrm{i}\omega t )$ convention, there is absorption if $\Im  (C_i) >0$ and gain if $\Im  (C_i) <0$.
The overall behavior of the acoustical system can be described by the transfer matrix $\mathsf{M}$
\begin{equation}
\left( \begin{array}{c} {p_2^+}\\ {p_2^-} \end{array} \right) = 
\left[ \begin{array}{cc} M_{11} &M_{12}\\ M_{21}  &M_{22} \end{array} \right]
\left( \begin{array}{c} {p_1^+}\\ {p_1^-} \end{array} \right)
\label{eq:1}
\end{equation}
where $p^+_{1,2}$ and $p^-_{1,2}$ are defined in Fig.  \ref{fig_1}(b).
After some algebra, the component of the overall transmission matrix are found to be:
\begin{eqnarray}
M_{11}&=& -\mathrm{i} \sin( k D) \frac{C_1\,C_2}{2}+\mathrm{e}^{\mathrm{i} k D} \left( 1+\frac{\mathrm{i}C_1}{2}+\frac{\mathrm{i}C_2}{2} \right)\nonumber\\
M_{12}&=& \mathrm{i} \sin( k D) \frac{C_1\,C_2}{2} -\mathrm{e}^{\mathrm{i} k D}\frac{\mathrm{i}C_1}{2} -\mathrm{e}^{-\mathrm{i} k D}\frac{\mathrm{i}C_2}{2}\nonumber\label{eq:MatM}\\
M_{21}&=& -\mathrm{i} \sin( k D) \frac{C_1\,C_2}{2}+\mathrm{e}^{-\mathrm{i} k D}\frac{\mathrm{i}C_1}{2}+\mathrm{e}^{\mathrm{i} k D}\frac{\mathrm{i}C_2}{2}\\
M_{22}&=& \mathrm{i} \sin( k D) \frac{C_1\,C_2}{2}+\mathrm{e}^{-\mathrm{i} k D} \left( 1-\frac{\mathrm{i}C_1}{2}-\frac{\mathrm{i}C_2}{2} \right)\nonumber
\end{eqnarray}
where in the case of a $\mathcal{PT}$ symmetric system \cite{stone2012}: $M_{11}=M^*_{22}$ and
$\Re[M_{12}] = \Re[M_{21}] = 0$. The transmission and reflection coefficients for waves coming from left and right are defined by
\begin{eqnarray*}
t_L= \frac{\mathrm{det}(\mathsf{M}) }{M_{22}} \:\:\:& &\:\:\: r_R=\frac{M_{12}}{M_{22}}\\
r_L= - \frac{M_{21}}{M_{22}} \:\:\:& &\:\:\: t_R= \frac{1}{M_{22}}
\end{eqnarray*}
Due to reciprocity we have  $\mathrm{det}(\mathsf{M})=1$ and then $t=t_L=t_R$.
As discussed in detail in \cite{stone2012},  by permutation of the outgoing waves, two different scattering matrices 
with different sets of eigenvalues can be defined, leading to distinct symmetry breaking. 
These two scattering matrices are 
\begin{equation}
\mathsf{S_r} = \begin{bmatrix} r_L & t\\t&r_R\end{bmatrix}
\,\,\,\text{and}\,\,\,
\mathsf{S_t} = \begin{bmatrix} t& r_L \\r_R&t\end{bmatrix}
\end{equation}

\begin{equation}
\text{where}  \begin{pmatrix} p_1^-\\ p_2^+ \end{pmatrix} 
 =\mathsf{S_r} \begin{pmatrix} {p_1^+}\\ {p_2^-} \end{pmatrix} , \,\,\,
\,\,\,\text{and}\,\,\,
\mathsf{S_t}=\mathsf{S_r} \mathsf{\sigma_x},
\end{equation}
$\mathsf{\sigma_x}$ is one of the Pauli matrices. 
The eigenvalues of $\mathsf{S_r}$ and  $\mathsf{S_t}$ may have both an exact and broken phases but the symmetry-breaking points are not the same. 
In this paper, we have chosen to consider both $\mathsf{S_r}$ and  $\mathsf{S_t}$ and the different phase transitions they imply.
When computing the scattering eigenvalues, it is useful to remind the $\mathcal{PT}$-symmetry conservation relations \cite{stone2011,stone2012,schomerus2013,mostafazadeh2014} that can be written for instance as $\mathsf{S_t^*}=\mathsf{S_t^{-1}}$ and leads to
\begin{eqnarray}
r^*_L r_R&=& 1-|t|^2\label{Eq_S1}\\
r_L t^* + r^*_L t&=&0\label{Eq_S2}\\
r_R t^* + r^*_R t&=&0\label{Eq_S3}
\end{eqnarray}
The eigenvalues of the scattering matrix $\mathsf{S_t}$ are given by $\lambda_{1,2}=t\pm \sqrt{r_R r_L})=t \left(1 \pm \sqrt{1-|t|^{-2}}\right)$. Then if $|t|<1$, the modulus of the eigenvalues is equal to 1. The case $|t|=1$ corresponds to symmetry-breaking and  $|t|>1$ correspond to the $\mathcal{PT}$-broken phase.
The eigenvalues of the other scattering matrix $\mathsf{S_r}$ are given by $s_{1,2}=(r_R+r_L \pm \sqrt{\Delta})/2$ where $\Delta = (r_R-r_L)^2+4t^2$. 
The broken phase condition can be written $\Delta=0$ which leads to $r_R-r_L=\pm 2 \mathrm{i} \,t$. In term of the transmission matrix coefficients, it is equivalent to $M_{12}-M_{21} = \pm 2 \mathrm{i}$ or $\Im(C_1) \sin( k D) = \pm 1$.

\begin{figure}[h]
\includegraphics[width=\columnwidth]{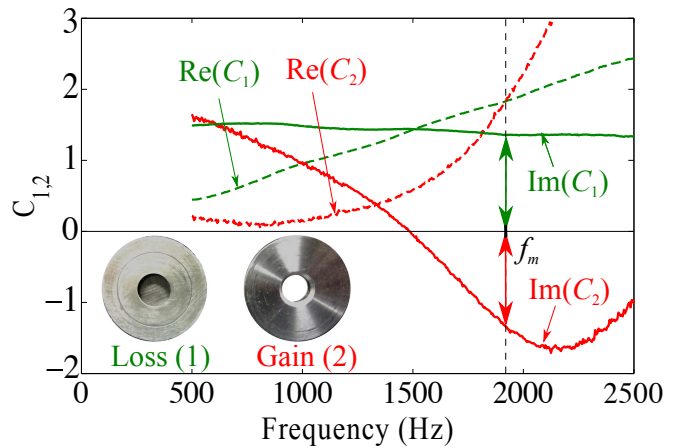}%
\caption{\label{fig_2} Real and imaginary part of the measured impedance parameters $C_1$ and $C_2$. 
When the imaginary part of $C_{ 2}$ is negative the diaphragm get some gain. At the frequency $f=f_m$, gain and loss are balanced $C_2=C_1^*$.}%
\end{figure}

{\it Experimental set-up}.---
As described in Fig. \ref{fig_1}, the $\mathcal{PT}$ symmetric system is mounted in a rigid circular duct between two measurement sections, upstream and downstream. 
Each measurement section consists in a  hard walled steel duct (diameter 30 mm) where two microphones are mounted. 
Two acoustic sources on both sides of the system give two different acoustic states and the four elements of the scattering matrix (transmission and reflection coefficient on both directions) for plane waves can be evaluated. 
A more detail description of the measurement technique can be found in \cite{testud2009}. 
The desired gain scatterer is realized by a finely designed diaphragm submitted to a steady flow. In this geometry, a shear layer is formed on its upstream edge and the flow is contracted into a jet with an area smaller than the hole of the diaphragm, see Fig. \ref{fig_1}(c). 
This shear layer is very sensitive to any perturbations like an oscillation in the velocity due to the acoustic wave. 
The shear layer convects and amplifies these perturbations (see the marked zone in Fig. \ref{fig_1}(c) and a strong coupling between acoustic and flow occurs when the acoustical period is of the order of the time taken by the perturbations to go from the upstream edge of the diaphragm to the exit of the diaphragm. 
This corresponds to a Strouhal number of the order of $S_h=f t/U_d \sim 0.2$ \cite{testud2009, lacombe2013} 
where $f$ is the frequency of the acoustic perturbation, $t$ is the thickness of the diaphragm and $U_d$ is the mean velocity in the diaphragm $U_d=U_0 (A/a)^2$ with $U_0$ the mean velocity in the duct 
and $a$ the radius of the diaphragm (Fig. \ref{fig_1}(c)). 
Eventually, this gain diaphragm has been chosen with an internal radius $a$ = 10 mm and a thickness $t$ = 5 mm (see Fig. \ref{fig_1} and the inset in Fig. \ref{fig_2}). 
The other diaphragm, that has to be lossy, has been chosen with an internal radius $a$ = 12 mm and a thickness $t$ = 4.3 mm. 
Two resistive metallic tissues have been glued to produce the dissipation by viscous and turbulent effects.

In a first step, the scattering coefficients of the two diaphragms have been measured separately, 
allowing us to deduce the values of the impedance  $C_{1,2}$.
These parameters,
 that have to  verify $C_2=C_1^*$ to get a $\mathcal{PT}$-symmetric system, 
are plotted on Fig. \ref{fig_2}. With the chosen geometry and flow parameters, it can be observed that there is
a frequency $f_m$ where the desired equality ($C_2=C_1^*$) is achieved.
In a second step, the scattering matrix of the system composed by the two balanced diaphragms is measured. 
All the subsequently reported measurements are made at the frequency $f_m$ = 1920 Hz and at the Mach number $\mathit{ M\kern-.20em a} $= 0.01 for which $C_2=C_1^*= 1.83 - 1.36\mathrm{i} $, allowing the system to be $\mathcal{PT}$ symmetric. 
In order to be able to observe the symmetry breaking, the distance between the two diaphragms $D$ is varied 
from 312 mm to 417 mm by inserting 22 rigid metallic tubes of different lengths.  
The minimal distance is chosen to minimize the hydrodynamical interactions between the two diaphragms. 
The maximal $D$ is chosen to have points over half a wavelength at the measurement frequency with
a value of $kD/2 \pi$ approximately in the range $1.7$ -- $2.4$.
\begin{figure}[h]
\includegraphics[width=\columnwidth]{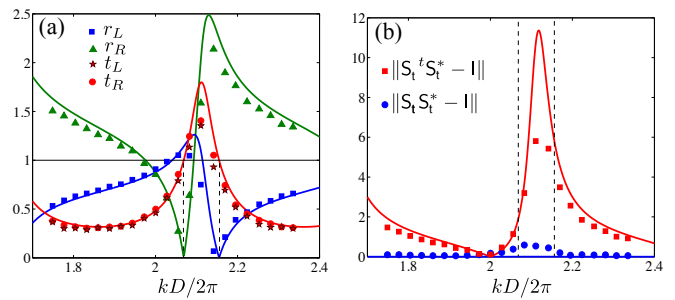}%
\caption{\label{fig_3} (a) Modulus of the  scattering coefficients.
(b) Norm of the deviation from energy conservation property and from the $\mathcal{PT}$-symmetry property
of the scattering matrix. 
Symbols correspond to experimental measurements and solid lines correspond to
the 1D theory.}
\end{figure}

{\it Results}.---
The measured transmission and reflection coefficients are displayed in Fig. \ref{fig_3}(a). 
They are compared to the theoretical values obtained by using the measured value of $C_1=C^*_2$ and the 1D modeling of Eqs. (\ref{eq:MatM}). 
The reflections from left $r_L$ (impinging on the loss) and right $r_R$ (impinging on the gain) appear as deeply asymmetric, with two points
with $|t|=1$ and $r_R=0$ or $r_L=0$. 
These two points correspond to the unidirectional transparency phenomenon where 
the wave passes unreflected with no amplitude change through the scatterers form one side, and is strongly reflected from the other side.
In order to verify experimentally the $\mathcal{PT}$ symmetry of the system,  in Fig. \ref{fig_3}(b), we plot the 2-norm of 
the matrix $ \mathsf{S_t} \mathsf{S_t^*} - \mathsf{I} $ corresponding to the shift from the  $\mathcal{PT}$ symmetry conservation relations in
Eqs \eqref{Eq_S1}-\eqref{Eq_S3}. 
For comparison the norm of the matrix $ \mathsf{S_t} ^t\mathsf{S_t^*} - \mathsf{I} $  which represents the deviation to the energy conservation is also displayed. 
It appears that $  \| \mathsf{S_t} \mathsf{S_t^*} - \mathsf{I} \|$ is nearly equal to zero in the whole range of paramaters  which unambiguously demonstrates
 that the system is  $\mathcal{PT}$ symmetric; meanwhile $  \| \mathsf{S_t} ^t \mathsf{S_t^*} - \mathsf{I} \|$  can take large values confirming 
 that our system strongly violates conservation of energy.
It can be noticed that for $kD$ multiple of $\pi$, the system is simultaneously $\mathcal{PT}$-symmetric and conservative; 
it can be verified (see Eqs. \ref{eq:MatM}) that in these cases the scattering is only sensitive to the real part of the impedances $C_1$ and $C_2$
ignoring thus the effect of gain and loss.
\begin{figure}[h]
\includegraphics[width=\columnwidth]{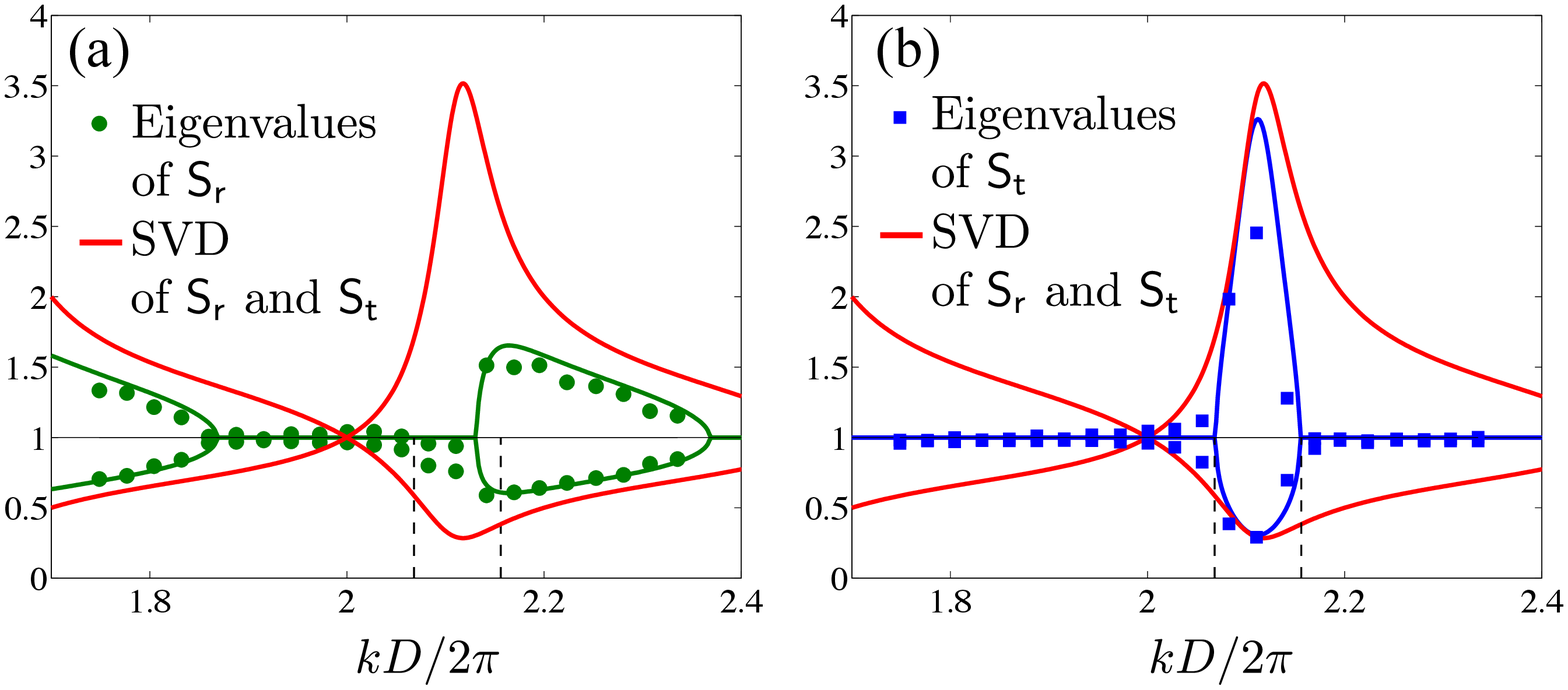}%
\caption{\label{fig_4} Spontaneous symmetry breaking of the scattering matrices. (a) green points: measurements, green line: 1D theory; 
(b) blue points: measurements, blue line: 1D theory. In each plot, the red lines corresponds the two SVD of the scattering matrix. 
Dashed lines correspond to $|t|=1$, i.e. phase transition for  $\mathsf{S_t}$.}%
\end{figure}

By  varying the length of the duct between the two diaphragms, we can also inspect the spontaneous symmetry breaking
of the scattering matrix of the system  \cite{stone2011}. 
In Fig. \ref{fig_4}, we show the eigenvalues of $S_r$ and $S_t$ that, since they  are different, 
lead to different symmetric and broken phases \cite{stone2012}. 
We represent also the singular value decomposition (SVD) of the scattering matrices. These two SVD are identical for  $\mathsf{S_t}$
and $\mathsf{S_r}$ (since $\mathsf{S_t} ^t \mathsf{S_t^*}=\mathsf{S_r} ^t \mathsf{S_r^*}$) and correspond respectively to the maximum
and minimum outgoing wave for any incoming waves with unit flux; by definition they are upper and lower bound of the modulus of the eigenvalues, and thus
must be different from one to allow the broken phase.
For each choice of scattering matrix, the experimental measurements, very close to the theoretical predictions, 
display clear signatures of the spontaneous symmetry breaking
with different broken phases for $\mathsf{S_t}$ and $\mathsf{S_r}$. 
In the symmetric phase the eigenvalues of the scattering matrices
remain on the unit circle in the complex plane, and the symmetry breaking corresponds to pairs of non-unimodular scattering eigenvalues
i.e. where the moduli are the inverse of each other and different from 1.
To the best of our knowledge, it is the first experimental demonstration of the symmetry breaking of the scattering matrix 
for $\mathcal{PT}$-symmetric systems as proposed in \cite{stone2011}.
\begin{figure}[h]
\includegraphics[width=\columnwidth]{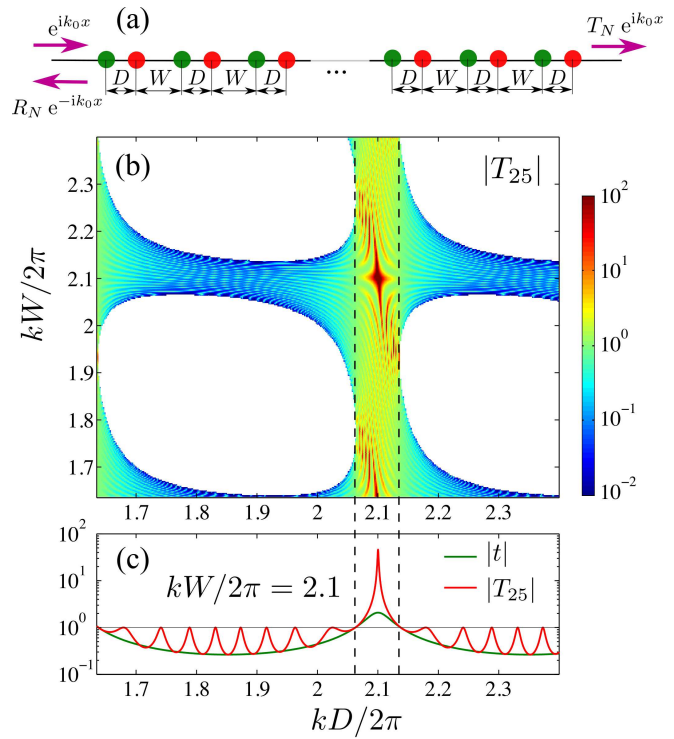}%
\caption{\label{fig_5} (a) Finite periodic case with $N$ cells, green(red) is a scatterer with loss (gain). (b) Transmission coefficient as a function of $kD$ and $kW$ for $N=25$. White regions correspond to band gaps. (c) Transmission for
$kW/ 2 \pi = 2.1$}.%
\end{figure}

In the broken phase, a particularly interesting case is the CPA-Laser where one eigenvalue of the $\mathsf{S}$ matrix goes to infinity (Laser) and the
other goes to zero (Absorber). From the experimental results of Fig. \ref{fig_4} we can see that this  Laser-Absorber is not achieved because the
maximum eigenvalue corresponds to  a $3.5$ amplification.
From Eqs. \eqref{eq:MatM}, it can be shown that the CPA-Laser condition can be obtained for larger values of the gain parameter ($\Im(C_2)\simeq2.5$) which cannot be achieved  with our current experimental setup. Nevertheless, in Fig. \ref{fig_5}, we show that quasi-CPA-Laser could be theoretically achieved by taking 
a finite periodic array of $N$ cells of our $\mathcal{PT}$-symmetric system with a  distance $W$ between each cell (Fig. \ref{fig_5}(a)).
The use of the 1D model  shows that very near CPA-Laser can be obtained by just tuning the number of cells and the intercell dimensionless frequency $kW$ ($N=25$ and $kW/2\pi=2.1$ in Fig \ref{fig_5}(b-c)). 
Fig. \ref{fig_5}(c) indicates that, by using interference Bragg effect in finite periodic case, it is possible to approach very closely the conditions of CPA-Laser.

{\it Conclusion}.--- 
Owing to vortex-sound interaction providing gain and loss in an acoustical system, 
we have obtained the experimental signatures of the spontaneous $\mathcal{PT}$-symmetry breaking in scattering systems. 
The scattering matrix eigenvalues can remain on the unit circle in the complex plane despite the Non-Hermiticity and the symmetry breaking results in pairs of scattering eigenvalues with inverse moduli.  The unidirectional transparency has also been observed.
It is noteworthy that this mechanical gain medium does not require to be electronically powered and that this $\mathcal{PT}$-symmetric system is very simple to manufacture: one tube, two diaphragms and a small flow inside the tube.
Therefore, this kind of acoustic system
can be seen as a building block to study wave propagation with more complex 
$\mathcal{PT}$-symmetry (for instance in periodic systems), 
and, more generally, we believe it provides an important connection 
between hydrodynamic instability theory, acoustic wave propagation and Non-Hermitian physics.

\end{document}